\documentclass[aps,pre,reprint,amsmath,amssymb]{revtex4-2}
\usepackage[pdftex]{graphicx}
\usepackage{subfigure}
\usepackage{dcolumn}
\usepackage{epstopdf}
\usepackage{color}
\usepackage{upgreek}
\usepackage[hidelinks]{hyperref}
\usepackage{ulem}

\newcommand{\D}[1]{\textcolor{black}{#1}}

\newcommand{\SD}[1]{\textcolor{black}{#1}} 

\makeatletter
\def\btt#1{\texttt{\@backslashchar#1}}%
\DeclareRobustCommand\bblash{\btt{\@backslashchar}}%
\makeatother

\begin{document}
\title{\D{Electrophoresis of metal-dielectric Janus particles with dipolar director symmetry in nematic liquid crystals}}

\date{\today}
\author{Dinesh Kumar Sahu,  and Surajit Dhara}
\email{Corresponding author: surajit@uohyd.ac.in} 
\affiliation{School of Physics, University of Hyderabad, Hyderabad 500 046, India}

\begin{abstract}

We study \D{electrophoresis} of metal-dielectric Janus particles with dipolar director symmetry in two nematic liquid crystals (LCs) having \D{same sign} of conductivity anisotropy ($\Delta\sigma$) but opposite \D{sign of} dielectric anisotropy ($\Delta\epsilon$). The applied ac electric field is parallel and perpendicular to the director for positive and negative dielectric anisotropy LCs, respectively. \D{We show that the Janus dipolar particles propel faster than the non-Janus dipolar particles in both LCs. The propelling speed of Janus dipolar particles is also significantly higher compared to the quadrupolar Janus particles studied previously.} We map the electroosmotic flow fields surrounding a \D{Janus dipolar} particle using the microparticle image velocimetry ($\upmu$-PIV) and show that the flow on metal hemisphere is stronger than that of dielectric hemisphere. \D{Altogether, Janus dipolar particles demonstrate efficient electrophoresis compared to both Janus and non-Janus quadrupolar particles. These findings may be useful for active matter, microrobotic and microfluidic devices. }
 
 \end{abstract}
\maketitle
\section{Introduction}

Creating active particles and studying their self-propelled or driven motility due to external force fields have been a subject of current interest~\cite{cb,jw,hr1,hr2,gl,dn,hrv}. The collective dynamics in dense systems of such particles are fundamentally important from the active matter perspectives~\cite{st,tp}.  These particles \SD{propel} at the expense of consumed energy and the system is far from  equilibrium~\cite{rama,rama1,aran,lau}. Usually, active particles are studied in isotropic \SD{medium} like water and the particles are mostly driven externally by light, electric or magnetic fields~\cite{cb}. The transport of particles by electric fields in fluids is known as electrophoresis and this phenomena has been utilised in many applications such as \SD{macro-molecular sorting, cargo transport, microrobots, display devices and colloidal assembly}~\cite{ramos,morgan,stv1,com,rc}. Electrophoresis is either linear or nonlinear depending on the charge and the shape polarity of the particles. In linear electrophoresis the \D{charged} particles move along the direction of applied electric field and the velocity varies linearly with field ($v\propto E$). In nonlinear electrophoresis or so called induced-charge electrophoresis the motility of particles is independent of the polarity of applied electric field and the propelling velocity is proportional to the square of the electric field ($v\propto E^2$)~\cite{sum,shen,taras,wu,murtsovkin,tm,baz1,baz2,baz3,baz4}.\\

The mobility of particles in nematic liquid crystals due to  electrophoresis is more intricate than that of its isotropic counter part and known as liquid crystal enabled electrophoresis (LCEEP)~\cite{oleg,od,oleg1}. Here, the electrophoresis is nonlinear and the mobility depends on the defect structure, in particular, the surrounding elastic distortions as well as the dielectric and conductivity anisotropies of  LCs~\cite{oleg2,pis,sag1,sag2}.
It has been shown that \SD{spherical silica particles inducing} quadrupolar director structure, namely particles with Saturn-ring and boojum defects are non-motile due to fore-aft symmetry of surrounding electroosmotic flows~\cite{oleg3}. This symmetry is however broken for a particle with point defect (dipolar particle) and  it can move along the director ${\bf\hat n}$  (direction of average molecular orientation). Very recently we have shown that the metal-dielectric Janus particles with quadrupolar director structure can be transported and the direction of motion is controllable by changing the amplitude and  frequency of the applied field~\cite{sd,sd1,sd2}.  

\begin{figure}[!ht]
\centering
\includegraphics[scale=0.26]{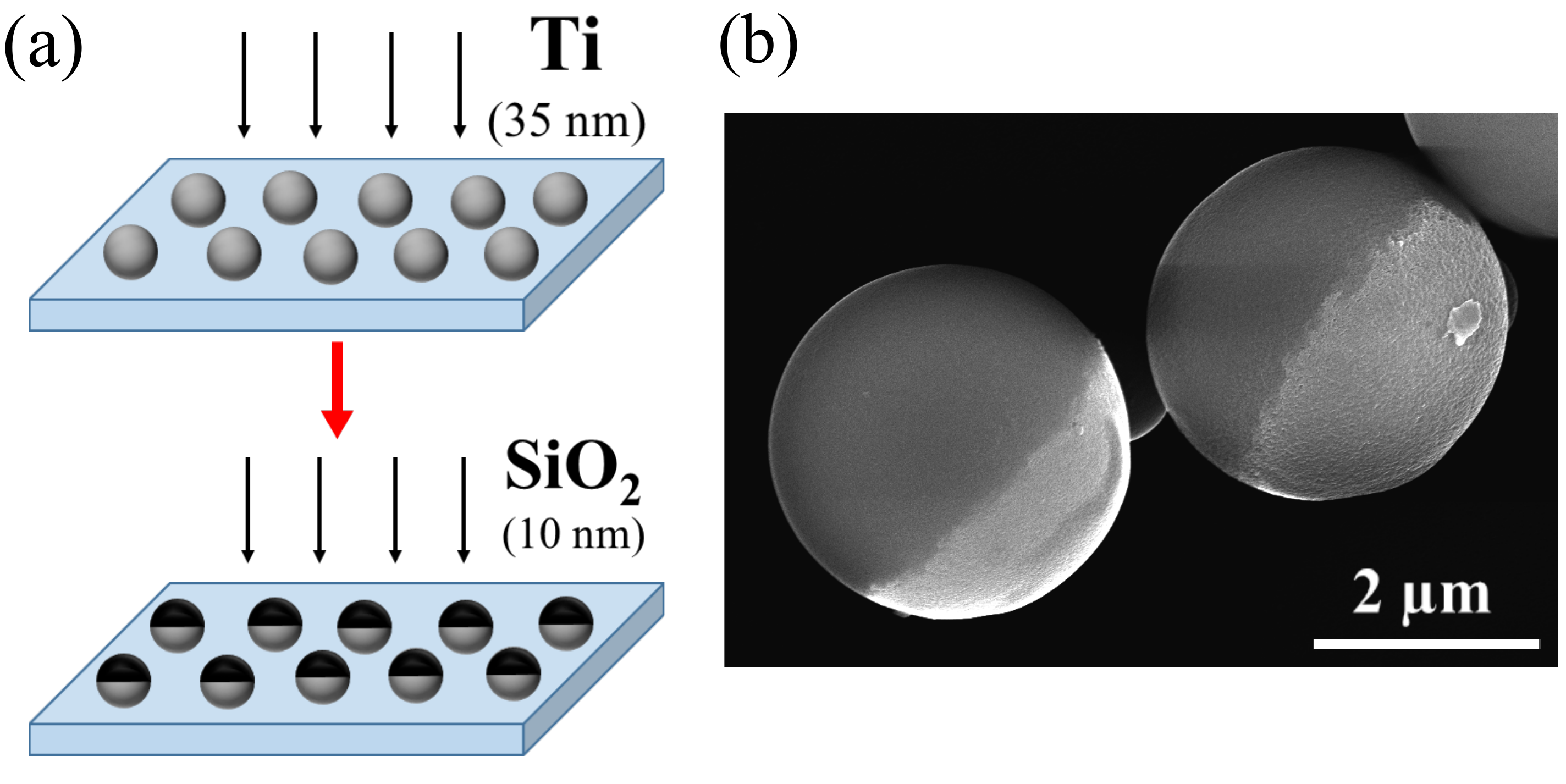}

\caption{({\bf a}) Coating method of Titanium (Ti) using electron beam deposition on silica microspheres. ({\bf b}) Scanning electron microscope (SEM) image of Janus particles.  The brighter (darker) region represents the metal (silica) hemisphere. }
\label{fig:figure1}
\end{figure}

In this paper we report experimental studies on the mobility of dipolar Janus particles in two nematic liquid crystals with positive and negative dielectric anisotropies. We show that the mobility of Janus particles in both nematic LCs is much higher than that of non-Janus particles. We map the surrounding flow fields using $\upmu$-PIV technique and show that not only the fore-aft symmetry of the flow field is broken but also the flow on the metal hemisphere is stronger than that on the dielectric hemisphere.

\section{Experiment}

Metal-dielectric Janus particles are prepared using normal deposition of Titanium (Ti) onto dry silica particles (SiO$_{2}$) of diameter $2a = 3.0 \pm 0.2$ $\upmu$m (Bangs Laboratories, USA)~\cite{stv1}. At first, the surface of a glass slide is made hydrophilic by treating with Piranha solution (Conc. H$_{2}$SO$_{4}$ : H$_{2}$O$_{2}$ :: 1 : 3) for 3-4 hours and then sonicated and rinsed with  distilled water.  Approximately $0.1-1$ wt$\%$ solution of microspheres are spread over an inclined ($45^{0}$) glass slide such that the solution flows down the slide leaving a monolayer of particles. After drying the slide, a thin layer ($\sim$35 nm) of Ti is coated on the monolayer by an electron beam deposition method such that only the exposed upper hemisphere of particles get coated (Fig.\ref{fig:figure1}(a)). Further, a thin layer ($\sim$10 nm) of SiO$_{2}$ is coated in order to make the surface of the particles chemically homogeneous. Then the particles are separated form the glass slide through ultrasonication for $1-2$ minutes and washed thoroughly. The scanning electron microscope (SEM) image of Janus particles is shown in Fig.\ref{fig:figure1}(b). \\

Finally, the surface of Janus particles is treated with N, N-dimetyl-N-octadecyl-$3$  aminopropyl-trimethoxysilyl chloride (DMOAP) in order to induce perpendicular (homeotropic) anchoring to the liquid crystal director~\cite{im,is}. A small quantity (0.1 wt$\%$) of DMOAP coated Janus particles is dispersed in nematic liquid crystals. We used two nematic liquid crystals, namely 5CB (pentyl cyanobiphenyl) and MLC-6608 (LC mixture obtained from Merck) and worked at room temperature.  \SD{The dielectric anisotropy $\Delta\epsilon$ of 5CB is positive whereas it is negative for MLC-6608 (Table-\ref{table1}). The conductivity anisotropy $\Delta\sigma$ for both LCs are positive.} For particle image velocimetry ($\upmu$-PIV) experiment we dispersed CdSe/ZnS quantum dots ($\sim$0.05 wt\%) in 5CB, using vortex mixer and ultrasonicator. \\

\begin{table}[ht!]
\large
\centering

\begin{tabular}{|c|c|c|c|c|}
\hline
Liquid crystals & $\Delta\epsilon$ & $\Delta\epsilon/\overline{\epsilon}$ & $\Delta\sigma/\overline{\sigma}$ & $\Delta\epsilon/\overline{\epsilon}$ - $\Delta\sigma/\overline{\sigma}$\\
\hline
5CB & 13 & 0.94 & 0.5 & 0.44\\
\hline
MLC-6608 & -3.8 & -0.64 & 0.9 & -1.54\\
\hline
\end{tabular}
\caption{\SD{Dielectric ($\Delta\epsilon=\epsilon_{||}-\epsilon_{\perp}$) and conductivity ($\Delta\sigma=\sigma_{||}-\sigma_{\perp}$) anisotropies of 5CB and MLC-6608 at room temperature. $\overline{\epsilon}=(\epsilon_{||}+\epsilon_{\perp}$)/2 and $\overline{\sigma}=(\sigma_{||}+\sigma_{\perp}$)/2.}}
\label{table1}
\end{table}

 We prepared Hele-Shaw type cells made of two parallel plates. Both upper and bottom plates were coated with AL-1254, cured at 180$^\circ$C for 1h and rubbed antiparallel way for obtaining planar alignment of the director.
For applying in-plane electric field  (parallel to ${\bf\hat n}$ for 5CB) the bottom plate was indium-tin-oxide (ITO) coated in which a small strip of width 2.0 mm is etched as shown schematically in Fig. \ref{fig:figure2}(a). For applying electric field orthogonal to ${\bf\hat n}$ (for MLC-6608) two ITO coated glass plates were used as shown in Fig. \ref{fig:figure4}(a). 
A function generator (Model: Tektronix AFG31000) and a voltage amplifier (Model: TEGAM 2350) were used for applying ac electric field to the cells.
 The mixture of Janus particles and LC is inserted into the cells via capillary action. An inverted optical polarising microscope (Ti-U, Nikon) was used for the experiments. The motion of the particles was  recorded using a 60$\times$ (NA=1.1) water immersion microscope objective (NIR Apo, Nikon) and a charge-coupled device (CCD) camera (iDs-UI) at the rate of 40 frames per second. The position of the particles was tracked from the recorded videos using a particle tracking program.

\section{Results and discussion}

\begin{figure}
\begin{center}
\includegraphics[scale=0.42]{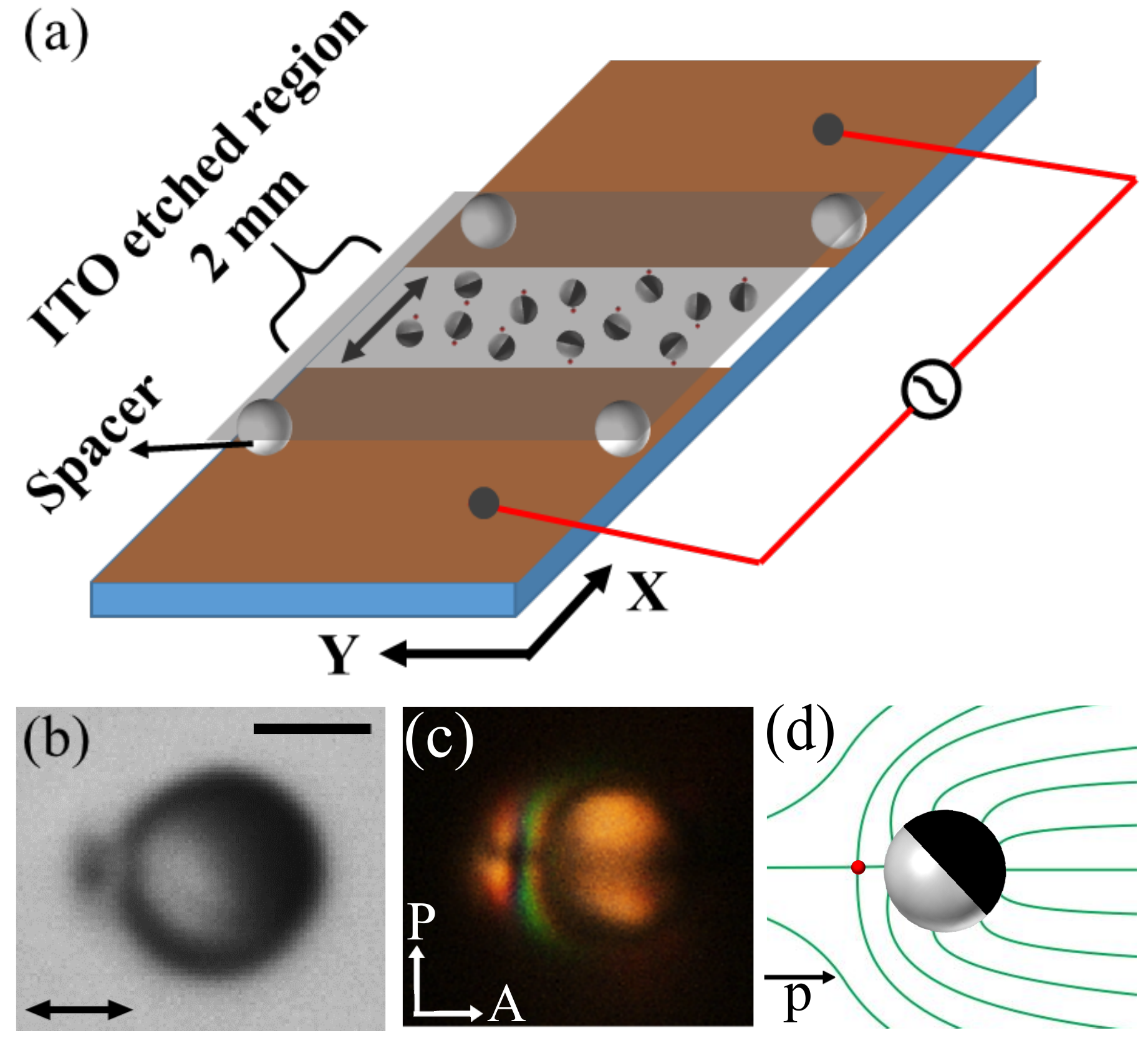}
\end{center}
\caption{ ({\bf a}) Diagram of a cell with in-plane electrodes. Double headed arrow indicates the director ${\bf\hat n}$, as well as the direction of the applied ac electric field. ({\bf b}) Optical micrograph and ({\bf c}) polarising optical micrograph of a dipolar Janus particle in 5CB. Double headed arrow indicates the director. Scale bar: 2 $\upmu$m. ({\bf d})  Dipolar director profile around a Janus  particle. \SD{$\vec{\mathrm{p}}$  defines the elastic dipole.}}
\label{fig:figure2}
\end{figure}

\begin{figure}[!ht]
\centering
\includegraphics[scale=0.35]{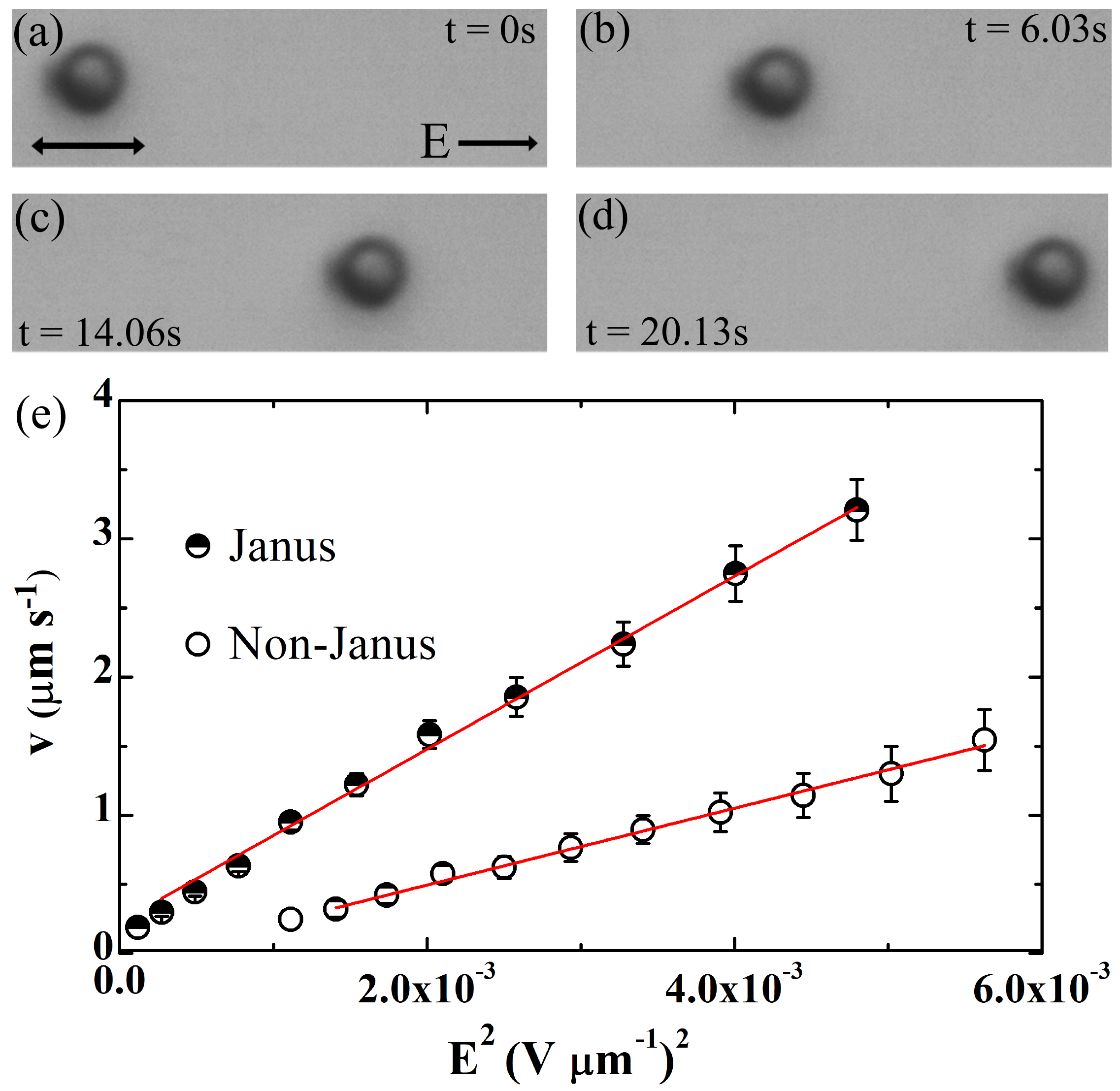}
\caption{({\bf a-d}) Sequence of images with elapse time showing mobility of a dipolar Janus particle along the director (see Movie S1)~\cite{sup}. Note that the direction of motion is parallel to the elastic dipole $\vec{\mathrm{p}}$. Double headed black arrow in (a) represents the director, ${\bf \hat{n}}$. ({\bf e}) Electric field dependent velocity ($v$) of a Janus and a non-Janus  particle. Red lines represent linear fits to $v\propto E^2$ with slopes 625 and 278 $\upmu$m\textsuperscript{3} s\textsuperscript{-1}V\textsuperscript{-2} for the Janus and the non-Janus particles, respectively. Error bars show the standard deviation of the mean value. }  
\label{fig:figure3}
\end{figure}

The DMOAP coated Janus particles \SD{nucleate} a hyperbolic hedgehog defect  of strength, s=-1 in nematic liquid crystal and the resulting director profile is dipolar as shown in Fig.\ref{fig:figure2}(d).  \SD{The elastic dipole $\vec{\mathrm{p}}$ points from -1 defect towards the centre of the particle.} A bright field image of a dipolar Janus particle in 5CB is shown in Fig.\ref{fig:figure2}(b). In transmitted light the metal coated hemisphere appears darker. The corresponding cross polarised image is also shown in Fig.\ref{fig:figure2}(c). 
When an ac electric field is applied, above a threshold field, the particles start propelling along the director, \D{irrespective of the orientation of the metal-dielectric interface. \SD{Here the electric field is applied parallel to the director ${\bf \hat{n}}$, in order to avoid elastic distortion due to the Freedericksz transition}. The direction of motion of particles is found to be} parallel to the elastic dipole $\vec{\mathrm{p}}$  (see Figs. \ref{fig:figure3}(a-d) and Movie S1)~\cite{sup}. This observation is very  different compared to multidirectional mobility of quadrupolar Janus particles in which the mobility is not restricted along the director. In particular, the direction of motion of quadrupolar particles depends on the orientation of the metal hemisphere with respect to the director ${\bf \hat{n}}$~\cite{sd,sd2}.

 Figure \ref{fig:figure3}(e) shows that the velocity of Janus particles varies quadratically with field as expected in liquid crystal enabled electrophoresis (LCEEP)~\cite{oleg}.  For comparative study, we have also present the field dependent velocity of a non-Janus dipolar particle. We observe that Janus particles \SD{propel} faster than non-Janus particles (Fig. \ref{fig:figure3}(e)). For example, at E\textsuperscript{2}$=4\times 10^{-3}$ \D{(V $\upmu$m\textsuperscript{-1})}\textsuperscript{2}, the velocities of the Janus and non-Janus particles are v\textsubscript{J}=2.8 $\upmu$m \D{s\textsuperscript{-1}} and v=1.03 $\upmu$m \D{s\textsuperscript{-1}}, respectively. It is also observed that the threshold field (above which the particles start propelling) for the Janus particles is less compared to the non-Janus particles. For example, threshold fields for the Janus and non-Janus particles are measured as E\textsubscript{J}$=1.1\times 10^{-4}$ \D{V $\upmu$m\textsuperscript{-1}} and E $=1.1\times 10^{-3}$ \D{V $\upmu$m\textsuperscript{-1}}, respectively.
 
 The electroosmotic velocity of LCs \D{around a particle} varies quadratically with electric field and it can be expressed as~\cite{oleg3,sat}:
 \begin{equation}
 u=CE^2.	
 \end{equation}
 The constant $C$ is given by
 \begin{equation}
C = \alpha \dfrac{\epsilon_{0}\overline{\epsilon}R}{\eta} \left( \dfrac{\bigtriangleup\epsilon}{\overline{\epsilon}}-\dfrac{\bigtriangleup\sigma}{\overline{\sigma}} \right),     
\end{equation}
 
 \noindent \SD{where $R$ is the radius of the particle and $\eta $ is the average viscosity of LC}~\cite{oleg3}.
 Here \D{the coefficient $\alpha\simeq1$ is introduced to account for the approximations such as 1/R as a measure of director gradients.}  The sign of the quantity in the parenthesis determines the direction of motion of  particles with respect to the elastic dipole $\vec{\mathrm{p}}$ ~\cite{sat}. If this quantity is \SD{positive}, the direction of motion is parallel to $\vec{\mathrm{p}}$ and if it is \SD{negative}, the direction of motion is antiparallel to $\vec{\mathrm{p}}$. \SD{For 5CB, $\Delta\epsilon/\overline{\epsilon}- \Delta\sigma/\overline{\sigma}=$ 0.44 \D{(Table-\ref{table1})} hence, the direction of motion is parallel to $\vec{\mathrm{p}}$.}\\
  
\begin{figure}[!ht]
\begin{center}
\includegraphics[scale=0.35]{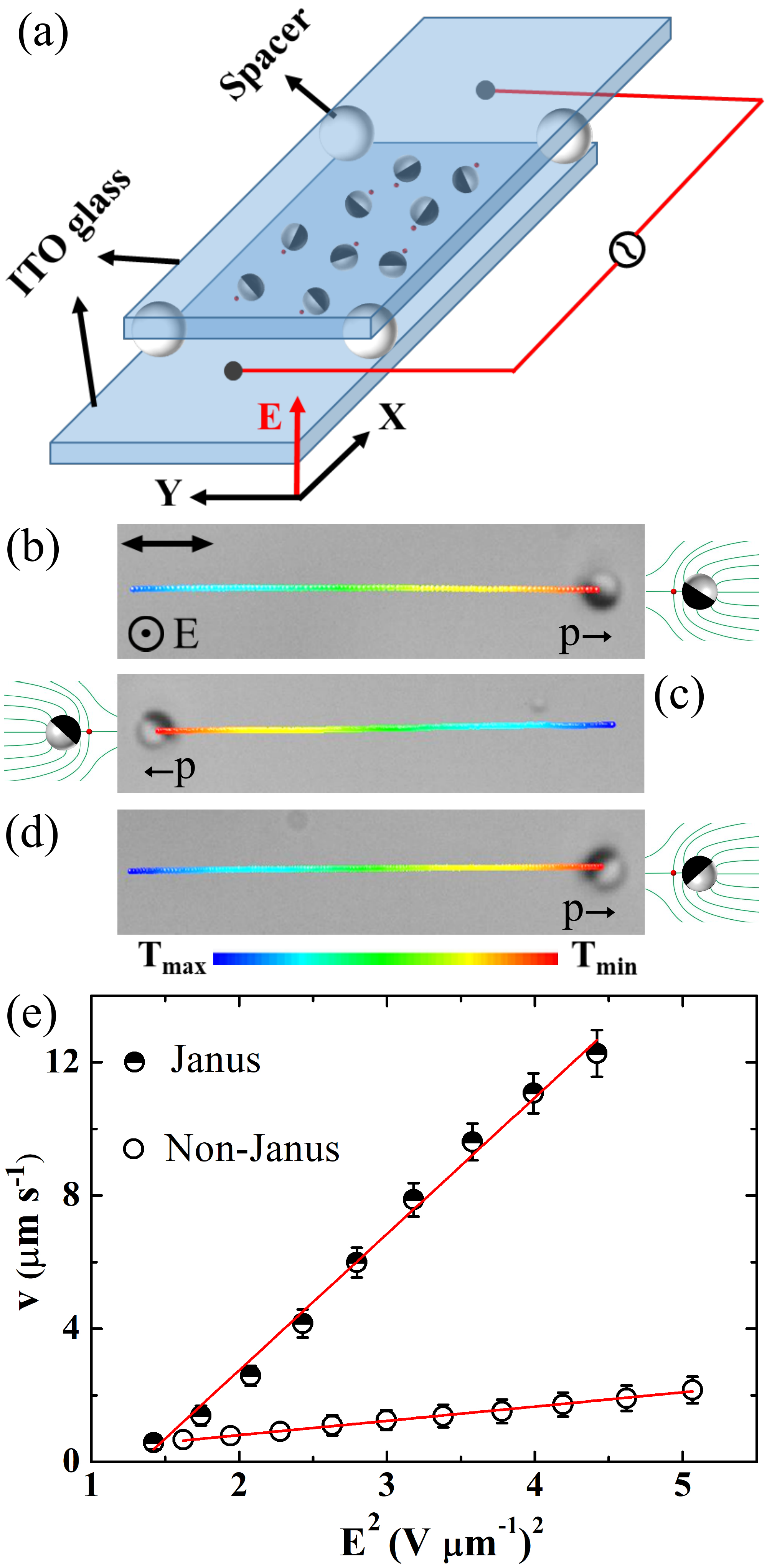}
\end{center}
\caption{({\bf a}) Diagram of a cell for applying electric field orthogonal to the director ${\bf \hat{n}}$. {\bf E} is along the z-axis  and the director is along the x-axis. ({\bf b-d}) Time coded trajectories of a few Janus dipolar particles in MLC-6608 with different orientations of the metal hemisphere under field amplitude of 2.0 V$\upmu$m$^{-1}$ at 30 Hz (see Movie S2)~\cite{sup}. T$_{\text{max}} = 5$ s, T$_{\text{min}} = 0$ s. Note that the direction of motion is antiparallel to elastic dipole $\vec{\mathrm{p}}$. ({\bf d}) Electric field dependent velocity ($v$) of a Janus and non-Janus dipolar particle. Red lines represent linear fits to $v\propto E^2$ with slopes 4.1 and 0.4 $\upmu$m\textsuperscript{3} s\textsuperscript{-1}V\textsuperscript{-2} for Janus and non-Janus particles, respectively. }
\label{fig:figure4}
\end{figure}

\begin{figure*}[!ht]
\begin{center}
\includegraphics[scale=0.35]{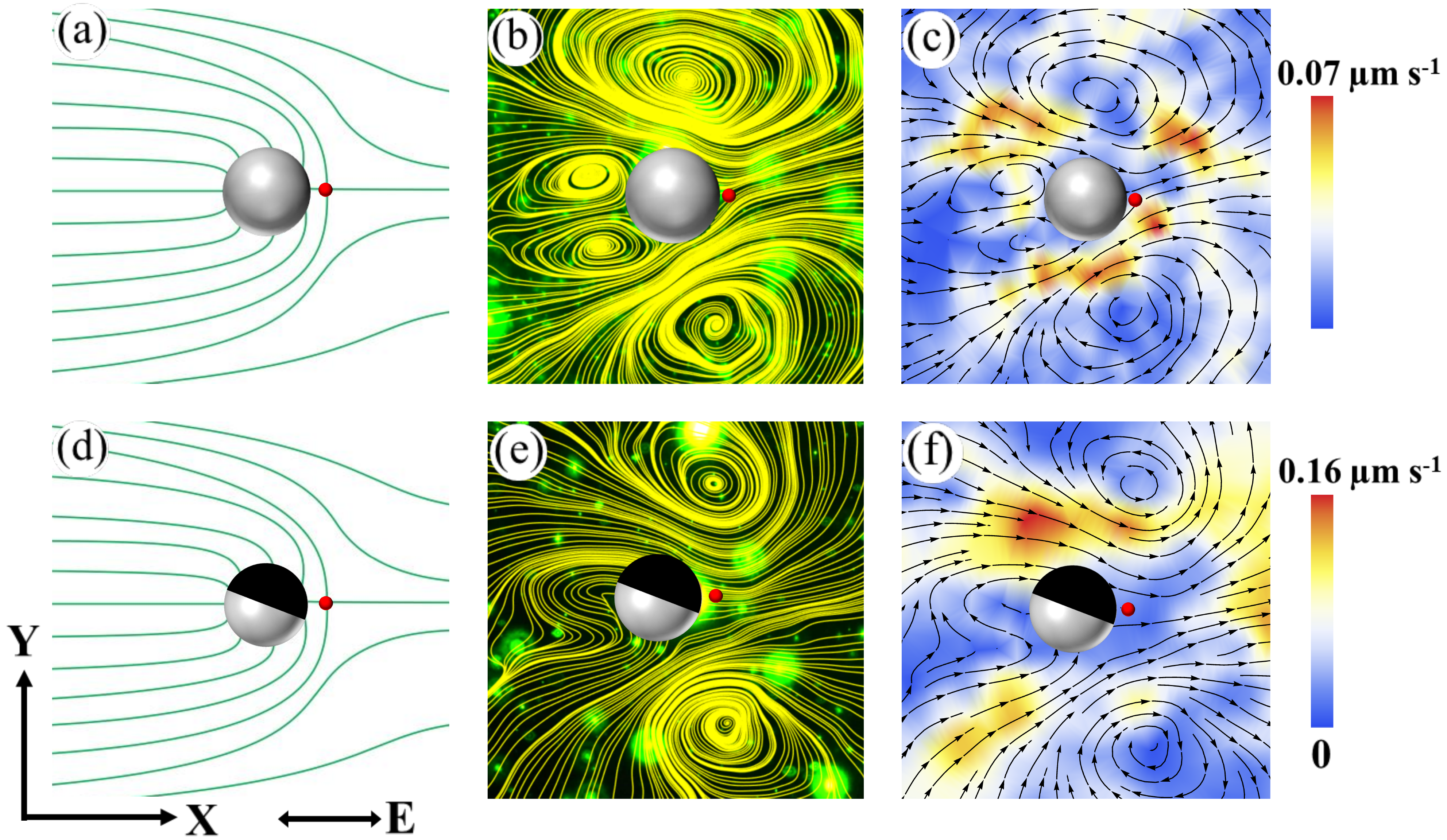}
\end{center}
\caption{({\bf a,d}) Director profile around a Janus and non-Janus particles. The applied electric field is parallel to the director i.e., along the x-axis. ({\bf b}) Streamlines of LC flow around a dipolar non-Janus particle. ({\bf c}) Directional streamlines i.e., the velocity map of the LCEO flows around the non-Janus dipolar particle.   ({\bf e}) Streamlines of LC flow  around a dipolar Janus particle. ({\bf f}) The velocity map of the LCEO flow around the Janus dipolar particle. \D{Red dots near the particle indicate the point defect.} Field amplitude: 9.0 mV$\upmu$m$^{-1}$, frequency: 30 Hz.}
\label{fig:figure5}
\end{figure*}

 In what follows we study \SD{electrophoresis} of Janus dipolar particles in MLC$-6608$. It exhibits negative dielectric anisotropy ( see Table-\ref{table1}).  In this sample, the electric field was applied perpendicular to the director ${\bf \hat{n}}$ \D{(Fig. \ref{fig:figure4}(a)) in order to avoid elastic distortion due to the Freedericksz transition}.
    Figures \ref{fig:figure4}(b-d) show time coded trajectories of three Janus dipolar particles with different orientations under the action of an ac electric field (see Movies S2)~\cite{sup}. The particles propel along the director but \SD{ with the point defect leading the way in contrary to the particles in 5CB.} It means the direction of motion is antiparallel to the elastic dipole $\vec{\mathrm{p}}$.  This is expected as the quantity in the parenthesis of Eq.(2) i.e.,  $\bigtriangleup\epsilon/\overline{\epsilon}-\bigtriangleup\sigma/\overline{\sigma}$ (=-1.54) is negative (\D{Table-\ref{table1}}). 
    Figure \ref{fig:figure4}(e) shows that the velocity of both the particles vary quadratically with field but the slope for Janus particle is about 10 time higher than that of the non-Janus particle. Thus Janus particles propel much faster as compared to the non-Janus particles.  \SD{ It may be noted that in 5CB LC this factor is about 2.2 times (see Fig.\ref{fig:figure3}(e)). This difference is expected as the velocity depends on the quantity $\bigtriangleup\epsilon/\overline{\epsilon}-\bigtriangleup\sigma/\overline{\sigma}$, which is about 3.5 times larger for MLC-6608 than that of 5CB (\D{Table-\ref{table1}}). Moreover, the applied electric field in MLC-6608 is much higher than that in 5CB. }

   In both samples the velocity of Janus dipolar particles is much higher than the non-Janus particles and this can be quantitatively understood based on the electroosmotic flows surrounding the particles.          
 We have used $\upmu$-PIV technique to observe the  electroosmotic flows in 5CB LC~\cite{piv}. For this purpose, we have chosen a bigger  particle of diameter $50 \pm 4 \upmu$m and fixed it with a glue in the middle of the cell.  Gap between the two in-plane electrodes is kept at 4 mm. A small quantity (0.01wt\%) of CdSe/Zn fluorescent quantum dots  (QDs) of size 1-2 nm is dispersed in  5CB and used as tracer particle.  The absorption maximum of the QDs is 512 nm and the emission wavelength is in the range of 530-550 nm. In the presence of  ac field QDs follow the  streamlines of electro-osmotic flows. \D{ Movies} of the flow are recorded using a Ni-S2 (Nikon) colour camera exposed at $300$ ms with a rate of one frame per second. The recorded file is saved as an array of images and analysed using $\upmu$-PIV software~\cite{piv}. The final flow pattern is obtained after averaging over nearly 100 images.   

Figures \ref{fig:figure5} (b) and (e) show the flow \D{streamlines} around a non-Janus and a Janus dipolar particle, \D{respectively}. \SD{In case of non-Janus particle, two bigger vortices are formed adjacent to the point defect and the smaller vortices are formed on the opposite side, consequently the fore-aft symmetry of the electroosmotic flow is broken.} Corresponding directional flow (Fig.\ref{fig:figure5}(c)) shows that the flow is outward along the director, similar to that of ``Pusher'' type micro-swimmers~\cite{oleg3}. The flow field of the Janus particle is somewhat different. \SD{Firstly, two small vortices on the left side of the microsphere disappear.}  Secondly, the velocity of fluid near the metal hemisphere is much higher than that on the silica hemisphere (see colour coded bars). For example, the maximum electroosmotic velocity for non-Janus particle is 0.07 $\upmu$m\D{s\textsuperscript{-1}}. Using Eq.(2) and considering $\overline{\eta}_{LC}\approx 90$ mPas, E=9 mV $\upmu$m\textsuperscript{-1}, R=25 $\upmu$m and taking $\Delta\epsilon$ and $\Delta\sigma$ from \D{Table-\ref{table1}}, we estimate the electroosmotic flow velocity for the non-Janus particle $u_{max}\approx 0.05~\upmu$m s\textsuperscript{-1}. This is very close to the value measured in the experiments (Fig.\ref{fig:figure5}(c)). 
\SD{In contrast} the maximum velocity for the Janus particle is 0.16 $\upmu$m \D{s\textsuperscript{-1}}, i.e., nearly double than than for non-Janus particle (Fig.\ref{fig:figure5}(f)). This is expected due to the higher polarisability of the metal hemisphere  than the dielectric hemisphere, consequently the higher induced charge density on the \SD{metal hemisphere}.  
    \begin{figure}[!ht]
\begin{center}
\includegraphics[scale=0.29]{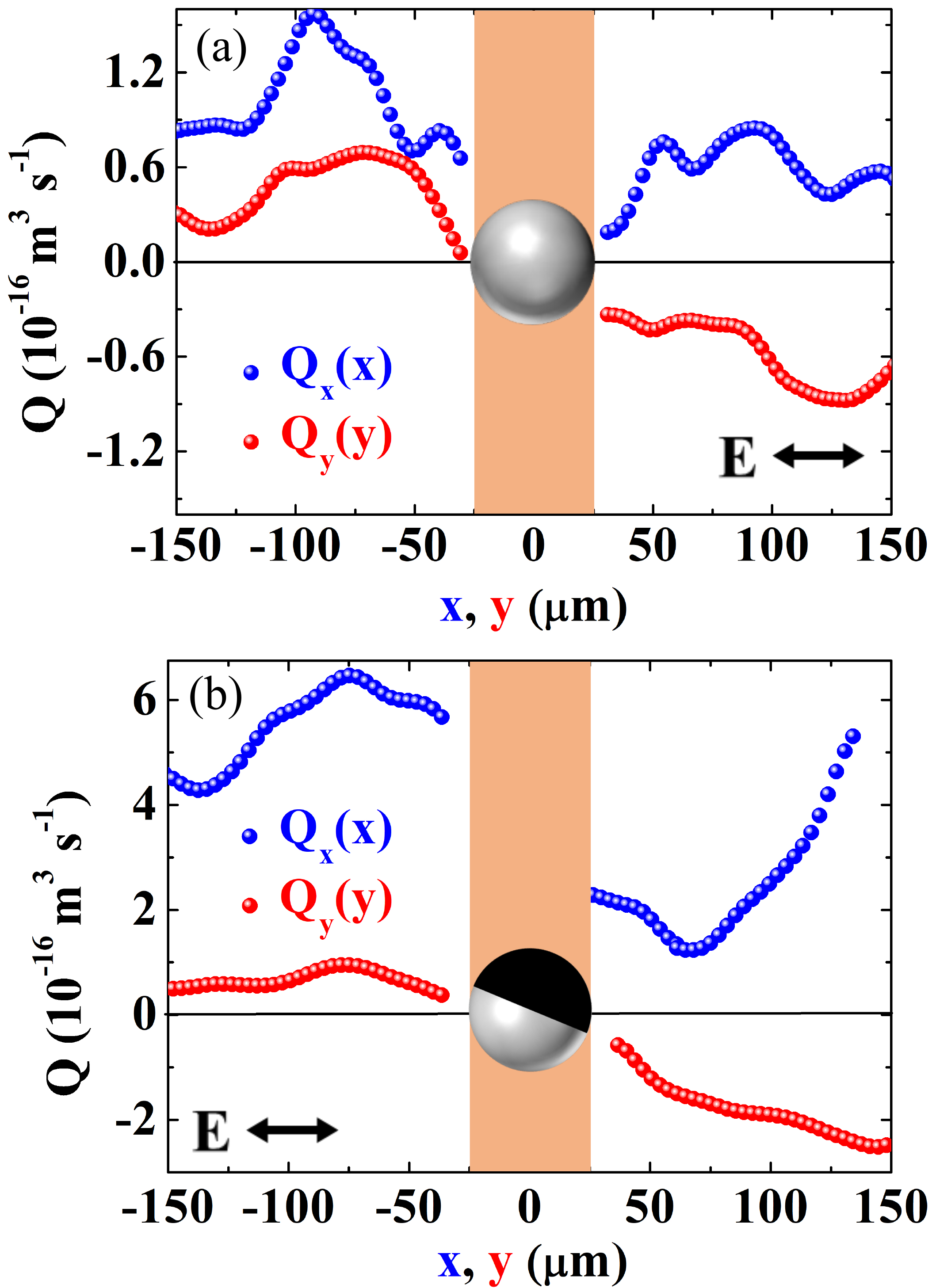}
\end{center}
\caption{ Volumetric flow functions $Q_{x}(x)$ and $Q_{y}(y)$ for ({\bf a}) non-Janus  and ({\bf b}) Janus particles. Pumping of LC in both cases is from the left to the right direction. \SD{Note that overall magnitude $Q_{x}(x)$ for Janus particle is larger than non-Janus particle.} }
\label{fig:figure6}
\end{figure}
 
We have calculated volumetric flows $Q_{x}(x)$ and $Q_{y}(y)$, \D{defined as the volume of fluid pumping per second around the sphere along  $x$ and $y$ directions of the cell, respectively. They can be expressed} as~\cite{oleg3}:
\begin{equation}
 Q_{x}(x)= \dfrac{2}{3}h\int^{y_{0}}_{-y_{0}} u_{x}(x,y)dy  
\end{equation}
\begin{equation}
Q_{y}(y)= \dfrac{2}{3}h\int^{x_{0}}_{-x_{0}} u_{y}(x,y)dx   
\end{equation}

\noindent where $u_{x}(x,y)$ and $u_{y}(x,y)$ are velocity components known from the experiments, $h$ is thickness of the cell and $x_{0}=y_{0}=150$ $\upmu$m. \SD{Calculated} $Q_{x}(x)$ and $Q_{y}(y)$ for both non-Janus and Janus particles are shown in Fig.\ref{fig:figure6}(a) and (b), respectively.
For non-Janus particle, $Q_x$ is positive on both left and right sides of the particle, that means the LC is pumped from left to right. On the contrary, $Q_y$ is positive on the left and negative on the right side of the particle, which means no net pumping of fluid along the y-direction. For Janus particle, the response is similar but the maximum $Q_x$ value is relatively larger than that of the non-Janus particle. It means that, comparatively the pumping of fluid from the left to the right is more for the Janus particle. \D{It may be mentioned that the magnitudes of  $Q_x$ and $Q_y$ depend on NLC's dielectric and conductivity anisotropies and  applied field strength. We obtained almost one order of magnitude less volumetric flow as our material is different and also our field value is almost 3 times smaller than that was used in Ref.~\cite{oleg3}.}

 \SD{
  Further, we make two observations in comparison to previous studies on quadrupolar Janus particles~\cite{sd,sd2}. Firstly, the threshold field required for Janus particles with hyperbolic hedgehog defect (E=1.2 V $\upmu$m\textsuperscript{-1}) is almost comparable to that of the particles with Saturn ring defects (E=1.3 V $\upmu$m\textsuperscript{-1}). But it is relatively lower compared to the particles with boojum defect (E=2.1 V $\upmu$m\textsuperscript{-1}). Secondly, the velocity of Janus dipolar particles is much higher compared to both quadrupolar Janus particles (i.e., with boojum and Saturn ring). For example, the velocity of a Janus dipolar particle at frequency 30 Hz and field $E$= 2.12 V $\upmu$m\textsuperscript{-1} is 12 $\upmu$m s\textsuperscript{-1}, whereas velocities for Janus particles with Saturn ring and boojum defects are 6.4 and 1.5 $\upmu$m s\textsuperscript{-1}, respectively. In particular, for a given field and frequency, velocity of Janus particles can be arranged in decreasing order as: V\textsubscript{hyperbolic hedgehog} $>$ V\textsubscript{Saturn ring}  $>$ V\textsubscript{boojum}. }

\section{Conclusion}

\SD{We have studied electrophoretic locomotion of Janus dipolar particles and} shown that the direction of motion of the particles is parallel to the  elastic dipoles in 5CB and antiparallel in MCL-6608, \D{respectively}. \SD{The velocity of the Janus particles is about 10 times higher than that of the non-Janus particles in 5CB and 2 times higher in MLC-6608.} 
We have mapped the electroosmotic flow fields in 5CB using $\upmu$-PIV \SD{and observed two vortices around the Janus particle as opposed to four vortices around the non-Janus particle.} The  flows on the metal hemisphere is stronger  and the volumetric flow of fluid along the direction of motion of Janus particles is \D{ nearly 2 times higher} compared to dielectric hemisphere. 
Our experiments demonstrate that the \D{propulsion of Janus dipolar particles in NLCs is highest compared to the both non-Janus and Janus quadrupolar particles. \D{For Janus quadrupolar particles, the surrounding medium (director profile) is symmetric and the asymmetric particle breaks the fore-aft symmetry of electroosmotic flows~\cite{sd,sd1}. In case of Janus dipolar particles, both the medium and the particles are asymmetric and they both are responsible for breaking symmetry of the flow.} Overall, the Janus dipolar particles display superior electrophoresis as far as propulsion is concerned, compared to both Janus and non-Janus quadrupolar particles in nematic liquid crystals.} Janus  particles with asymmetric shape, higher order elastic moment and genus could offer unusual motility and dynamics in liquid crystals.  \\

{\bf Acknowledgments:}
SD  acknowledges the support from the Department of Science and Technology, Govt. of India (DST/SJF/PSA-02/2014-2015),  DST-PURSE-II and UoH (UoH/IoE/RC1-20-010). DKS acknowledges DST INSPIRE fellowship.


\begin{thebibliography}{99}

\bibitem{cb} C. Bechinger, R. D. Leonardo, H. L{\"o}wen, C. Reichhardt, G. Volpe, and G. Volpe, \textit{Rev. Mod. Phys.}, 2016, {\bf 88}, 045006.

\bibitem{jw}J. Wang and W. Gao, \textit{ACS Nano}, 2012 {\bf 6}, 5745-5751.

\bibitem{hr1} H. R. Jiang,  H. Wada, N. Yoshinaga, and M. Sano, \textit{Phys. Rev. Lett.}, 2009, {\bf 102}, 208301.

\bibitem{hr2} H. R. Jiang, N. Yoshinaga, and M. Sano, \textit{Phys. Rev. Lett.}, 2010, {\bf 105}, 268302.

\bibitem{gl} G. Li and  J. X. Tang, \textit{Phys. Rev. Lett.}, 2009, {\bf 103}, 078101.

\bibitem{dn} D. Nishiguchi and M. Sano, \textit{Phys. Rev. E}, 2015, {\bf 92}, 052309.

\bibitem{hrv}H. R. Vutukuri, M. Lisicki, E. Lauga and J. Vermant, \textit{Nat. Commun.}, 2020, {\bf 11}, 2628.

\bibitem{st} S. Thutupalli, R. Seemann, and S. Herminghaus, \textit{New J. Phys.}, 2011, {\bf 13}, 073021.

\bibitem{tp} P. Tierno,  R. Golestanian, I. Pagonabarraga, and F. Sagu{\'e}s, \textit{Phys. Rev. Lett.}, 2008, {\bf 101}, 218304.

\bibitem{rama} S. Ramaswamy, \textit{Annu. Rev. Condens. Matter Phy.}, 2010, {\bf 1}, 323-345. 

\bibitem{rama1} M. C. Marchetti, J. F. Joanny, S. Ramaswamy, T. B. Liverpool, J. Prost, M. Rao, and R. A. Simha, \textit{Rev. Mod. Phy.}, 2013,  {\bf 85}, 1143.

\bibitem{aran} I. S. Aranson, \textit{Phys. Usp.}, 2013, {\bf 56}, 79.

\bibitem{lau} E. Lauga and T. R Powers, \textit{Rep. Prog. Phys.}, 2009, {\bf 72}, 096601. 

\bibitem{ramos} A. Ramos, \textit{Electrokinetics and Electrohydrodynamics in Microsystems} (Spinger, 2011).

\bibitem{morgan} H. Morgan and N. G. Green, \textit{AC Electrokinetics: Colloids and nanoparticles} (Research Studies Press Ltd, 2003).

\bibitem{stv1}J. Yan, M. Han, J. Zhang, C. Xu, E. Luijten, and S. Granick, \textit{Nat. Mater.}, 2016, {\bf 15}, 1095-1099.

\bibitem{com} B. Comiskey, J.D. Albert, H. Yoshizawa, and J. Jacobson, \textit{Nature}, 1998, \textbf{394}, 253-255.

\bibitem{rc} R. C. Hayward, D. A. Saville, and I. A. Aksay, \textit{Nature}, 2000, \textbf{404}, 56-59.

\bibitem{sum} S. Gangwal, O. J. Cayre, M. Z. Bazant, and O. D. Velev, \textit{Phys. Rev. Lett.}, 2008,  \textbf{100}, 058302.

\bibitem{shen} \D{C. Shen, Z. Jiang, L. Li, J. F. Gilchrist and H. D. Ou-Yang, \textit{Micromechanics},2020, {\bf 11}, 334.}

\bibitem{taras} \D{T. Y. Molotilin, V. Lobaskin and O. I. Vinogradova, \textit{J. Chem. Phys.}, 2016 {\bf 145}, 244704.}

\bibitem{wu} M. Fuduo, Y. Xingfu, Z. Hui, and N. Wu, \textit{Phys. Rev. Lett.}, 2015, {\bf 115}, 208302.

\bibitem{murtsovkin} V.A. Murtsovkin, \textit{Colloid Jour.}, 1996, \textbf{58}, 341. 

\bibitem{tm} T. M. Squires and S. R. Quake, \textit{Rev. Mod. Phys.}, 2005, \textbf{77}, 977.

\bibitem{baz1} M. Z. Bazant, M.S. Kilic, B.D. Storey, and A. Ajdari, \textit{Adv. Colloid Interface Sci.}, 2009, \textbf{152}, 48-88.

\bibitem{baz2} T. M. Squires and M. Z. Bazant, \textit{J. Fluid Mech.}, 2004, \textbf{509}, 217-252.

\bibitem{baz3} M.Z. Bazant and T. M. Squires, \textit{Phys. Rev. Lett.}, 2004, \textbf{92}, 066101.

\bibitem{baz4} T.M. Squires and M. Bazant, \textit{J. Fluid Mech.}, 2006, \textbf{560}, 65-101. 

\bibitem{oleg1} O. D. Lavrentovich, I. Lazo, and O. P. Pishnyak, \textit{Nature}, 2010, \textbf{467}, 947-950.

\bibitem{oleg} I. Lazo and O. D. Lavrentovich, \textit{Phil. Trans. Soc. A}, 2013, \textbf{371}, 2012255.

\bibitem{od} O. D. Lavrentovich, \textit{Curr. Opin. Colloid Interface Sci.}, 2016, \textbf{21}, 97-109.

\bibitem{oleg2}O. D. Lavrentovich, \textit{Soft Matter}, 2014, {\bf 10}, 1264-1283.

\bibitem{pis} O.P. Pishnyak, S. Tang, J. R. Kelly, S.V. Shiyanovskii, and O.D. Lavrentovich, \textit{Phys. Rev. Lett.}, 2007, \textbf{99}, 127802.

\bibitem{sag1} S. Hern{\`a}ndez-Navarro, P. Tierno, J. A. Farrera, J. Ign{\'e}s-Mullol, and F. Sagu{\'e}s, \textit{Angew. Chem. Int. Ed.}, 2014, \textbf{53}, 10696-10700.

\bibitem{sag2} A. V. Straube, J. M. Pag{\'e}s, P. Tierno, J. Ign{\'e}s-Mullol, and F. Sagu{\'e}s, \textit{Phys. Rev. Res.}, 2019, \textbf{1}, 022008(R).


\bibitem{oleg3} I. Lazo, C. Peng, J. Xiang, S. V. Shiyanovskii, and O. D. Lavrentovich, \textit{Nat. Commun.}, 2014, \textbf{5}, 5033.

\bibitem{sd} D. K. Sahu, S. Kole, S. Ramaswamy and S. Dhara, \textit{Phys. Rev. Res.}, 2020, \textbf{2}, 032009(R).

\bibitem{sd1} D. K. Sahu, and S. Dhara, \textit{Phys. Rev. Applied.}, 2020, \textbf{14}, 034004.

\bibitem{sd2} D. K. Sahu, and S. Dhara, \textit{Phys. Fluids}, 2021, {\bf 33}, 0187106.

\bibitem{im} I. Mu\v{s}evi\v{c}, \textit{Liquid crystal colloids}. 2017. (Springer Inter- national Publishing AG).

\bibitem{is} I. I. Smalyukh, \textit{Annu. Rev. Condens. Matter Phys.}, 2018, {\bf 9}, 207-226.

\bibitem{sup} Supplemental Information for materials data and description of movies.

\bibitem{sat}S. Paladugu, C. Conklin, J. Vi\~nals, and O. D. Lavrentovich, \textit{Phys. Rev. Appl.}, 2017, {\bf 7}, 034033.

\bibitem{piv} W. Thielicke and E. J. Stamhuis, Time Resolved Digital Particle Image Velocimetry Tool for MATLAB. http://pivlab.blogspot.com/p/pivlabdocumentation.html (2010).


\end{thebibliography}
\end{document}